\documentclass{aastex6}

\fullcollaborationName{The Friends of AASTeX Collaboration}

\begin{document}


\title{Study of Dark-Matter Admixed Neutron Stars using the\\ Equation of State from
the Rotational Curves of Galaxies}

\author{Z. Rezaei}
\affil{Physics Department
and Biruni Observatory \\
College of Sciences, Shiraz
University \\
Shiraz 71454, Iran}


\begin{abstract}
In this work, we employ the dark matter equations of state (DMEOSs) obtained
from the rotational curves of galaxies as well as the fermionic DMEOS with $m=1.0\ GeV$ to study the
structure of dark-matter admixed neutron stars (DMANSs). Applying the equation of state
 in the Skyrme framework
for the neutron matter (NM), we calculate the mass-radius relation for different DMANSs
with various DMEOSs and central pressure of dark matter (DM) to NM ratios.
Our results show that for some DMEOSs, the mass-radius relations are in agreement with
 new observations, e.g. EXO 1745-248, 4U 1608-52, and 4U 1820-30, which are inconsistent with the normal neutron stars.
We conclude that both DMEOSs and central pressure ratios of DM to NM
affect the slope of the mass-radius relation of DMANSs.
This is because of
the interaction between DM and NM, which leads to gravitationally or self-bound DMANSs.
We study the radius of the NM sphere as well as the radius of the DM halo for different DMANSs.
The results confirm that, in some cases, a NM sphere with a small radius is surrounded
by a halo of DM with a larger radius.
Our calculations verify that, due to the different degrees of
DM domination in DMANSs, with a value of the visible radius of a star
two possible DMANSs with different masses can be exist.
The gravitational redshift is also
calculated for DMANSs with different DMEOSs and central pressure ratios.
The results explain that the existence of DM in a DMANS leads to higher
values of gravitational redshift of the star.
\end{abstract}




\section{Introduction} \label{sec:intro}
A neutron star is a compact star which contains many neutrons.
The theory of general relativity predicts the existence of a maximum mass above which
the neutron star is unstable and collapses into a quark star or black hole.
The maximum mass of a neutron star depends on the interaction between particles, and thus
the equation of state (EOS) of the system.
The stiffer the EOS, the larger the maximum mass of neutron star.
In addition, the rotation of the neutron star leads to higher values
of the maximum mass. Since the neutron star contains dense asymmetric nuclear matter, one should apply the EOS
of strongly interacting asymmetric nuclear matter to investigate the bulk properties of the star.
In addition, some authors have studied the effects of the existence of other particles, such as hyperons, meson condensates, and quarks, in the core
of neutron stars on the maximum mass \citep{Ozel5,Alford,Zhou}.

From observation, radio pulsars show large masses of about $2\ M_{\odot}$
\citep{Ozel5,Demorest,Antoniadis}.
For example, according to \citep{Ozel5}, the mass of neutron star EXO 07482676 is measured to be $2.10\pm0.28M_{\odot}$ and the radius equal to $13.8\pm1.8\ km$.
From theory, stiff EOSs like the neutron matter (NM) EOS in SLy230a and SLy230b models
\citep{Chabanat}, can result in such large masses and  mass-radius relations.
Nevertheless, in some observations-such as EXO 1745-248 with mass $M = 1.4\pm0.1\ M_{\odot}$ and radius $R = 11\pm 1\ km$ \citep{Ozel},
the neutron star
in 4U 1608-52 with mass $M = 1.74\pm0.14\ M_{\odot}$  and radius  $R =  9.3 \pm 1.0\ km$ \citep{Guver2}, and also  the neutron star in
the low-mass X-ray binary 4U 1820-30 with mass $M = 1.58\pm0.06\ M_{\odot}$ and radius  $R = 9.1 \pm 0.4\ km$ \citep{Guver9},- the results are different.
 It can be shown that soft EOSs are needed to lead these mass-radius relations from observation.
 Different equations of state for pure nuclear matter that have been obtained
using different methods, such as the variational method with $AV_{18}$ potential plus the UIX potential
or $AV_{18}$ potential plus a three-body UIX potential
\citep{Akmal}, and Dirac-Brueckner-Hartree-Fock models \citep{Prakash,Engvik} cannot predict some observational
data, such as 4U 1608-52 and 4U 1820-30 \citep{Lattimer1,Lattimer}.
Therefore, this rules out the possibility of the existence of the pure nuclear matter in the neutron star.
One explanation for these observations is that the existence of some particles, such as quarks,
mesons, and hyperons, in the neutron star can lead to soft EOSs
and the mass-radius relations in agreement with these observations \citep{Weber5,Page,Alford,Klahn}.

In addition, another possibility that leads to these
mass-radius relations is the existence of dark matter (DM) in the neutron star.
Cosmological structure \citep{Springel}, gravitational lensing
\citep{Massey}, and galactic rotational curve \citep{Weber} are some observations that could confirm the existence of DM.
According to \citep{Ade}, the
contributions of ordinary matter, DM, and dark energy
in the universe are $4.9\%$, $26.8\%$, and $68.3\%$, respectively.
Supposing the existence of DM, it should be present in
all astrophysical objects \citep{Sandin}.
The DM may be accreted
to the neutron stars because of the large density of the neutron star matter \citep{Bertone,Angeles,Fuller}.
Therefore, to study models of DM,
neutron stars are of much interest in astrophysics and astroparticle physics \citep{Goldman,Bertone,Kouvaris12}.
It has been shown that self-annihilating neutralino WIMP
 DM accreted onto neutron stars may lead to seeding
 compact objects with long-lived lumps of strange-quark matter
\citep{Perez-Garcia}.
The heating caused by possible DM annihilation in neutron stars  may be an observable \citep{Lavallaz} since
the neutron star itself does not burn.
In addition, it has been indicated that self-annihilation
of DM in neutron stars can change their linear and angular momentum
\citep{Angeles}, as well as the cooling properties
of neutron stars \citep{Kouvaris}.
The critical phases
of stellar evolution could be modified in the presence of
relatively small amounts of DM \citep{Sandin}.

Because of the possibility of the existence of DM in neutron stars,
a model called the dark-matter admixed neutron star (DMANS) has been considered
 \citep{Sandin,Ciarcelluti,Leung11,Leung12,Xiang}.
The mechanisms for accumulating
DM in a neutron star can result both from the stellar
formation process and from subsequent accumulation by
accretion of DM particles during the stellar lifetime \citep{Ciarcelluti}.
The DM inside the stars alters the structure and may lead to the collapse of a
neutron star \citep{Kouvaris12}.
The structure of these stars is still an open issue because of the unknown nature
of DM. However, it is useful to study the structure of DMANSs using existing
information on DM.
The DMANS is a two-fluid system where
the neutron star matter and DM interact only
through gravitational force \citep{Leung12}.
In order to study the structure of these stars, the Tolman-Oppenheimer-Volkoff (TOV) equation is separated into two different
sets for the neutron star and dark components inside the star
\citep{Sandin,Lavallaz,Ciarcelluti,Leung11,Leung12}.

The mass-radius relation of DMANSs which is affected by the DM, is one of the observational
results that is possible to measure.
The basic equation for DM to study the structure of DMANSs is the DM EOS \citep{Sandin,Ciarcelluti,Leung11,Xiang}.
Applying a method that combines kinematic and gravitational lensing
data to test the widely adopted assumption of pressureless DM,
the DM EOS has been measured \citep{Serra1}.
Moreover, by modeling galactic halos describing the DM as a nonzero pressure
fluid and using observational data of the rotation curves of galaxies,
a DM EOS has been obtained \citep{Barranco}.

Some previous studies have been performed to investigate the structure of DMANSs
\citep{Sandin,Ciarcelluti,Leung11,Leung12,Li,Xiang}.
The general-relativistic hydrostatic equations have been generalized to spherical
objects with multiple fluids that interact by gravity \citep{Sandin}.
Moreover, assuming that the microphysics is the same in the two sectors of the DMANSs,
the effects of mirror DM on neutron star structure have been studied.
It was concluded that the structure depends on the relative
number of mirror baryons to ordinary baryons.
Supposing that DM is made of some form of stable and long-living particles that can accumulate in the star, it has been shown that
all mass-radius measurements can be explained with one nuclear matter EOS and a dark core of varying relative size \citep{Ciarcelluti}.
Consequently, observational data, which provide a test of the theory,
will become a powerful tool for the determination
of DM \citep{Ciarcelluti}.
Considering non-self-annihilating DM particles of mass $1.0\ GeV$
along with normal nuclear matter, and using a general relativistic two-fluid formalism,
the properties of DMANSs have been studied \citep{Leung11}.
It was found that a new class of compact stars
consisting of a small normal matter core with radius of a few km within a 10 km DM halo can exist.
Employing a general relativistic, two-fluid formalism to study the admixture
of degenerate DM and normal nuclear matter shows that
a new class of compact
stars that are dominated by DM can exist \citep{Leung12}. These stars
have a small neutron star matter core, with radius of a few km embedded
in a larger 10 km DM halo.
In addition, these DMANSs have two classes of oscillation modes.
The first class of modes reduces
to the same set of modes for ordinary neutron stars without DM, and the second
class of modes is due to DM.
In a consistent DMANS model, considering
DM particles to behave like fermions which interact with a certain
repulsive interaction, it has been found that DM would soften the
EOS more strongly than hyperons, reducing the maximum
mass of the star \citep{Li}. However, with small mass particles, the
maximum mass could be larger than $2\ M_{\odot}$.
The effects of fermionic DM on the properties of neutron stars using the two-fluid TOV formalism have been studied \citep{Xiang}. It has been shown that
the mass of DM candidates, the amount of DM, and interactions among DM candidates
affect the mass-radius relationship. In addition, the DM in a DMANS results in a spread of
mass-radius relationships, and in some
cases the DM distribution can surpass the NM distribution to form a DM halo.
It has been confirmed that the DM admixture in neutron stars leads to the
shrinkage of the NM surface, which results in the observation of a small radius.
Moreover, the DM distribution can surpass the
neutron star matter distribution and form a DM halo when the DM candidates have low mass or
there is repulsion of DM. With a the large DM
fraction in the DMANSs, the repulsive interaction among DM may result in stars with a mass of
above $2\ M_{\odot}$.
In the present work, we investigate the properties of DMANSs using the Skyrme interaction
in the neutron star matter and the DM EOS obtained
from the rotational curves of galaxies in a general relativistic, two-fluid formalism.

\section{Formalism} \label{sec:style}
	
\begin{figure*}
\includegraphics{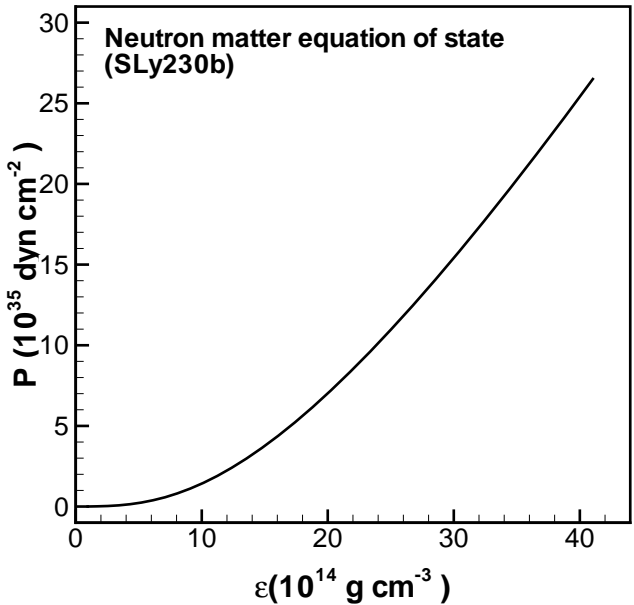}
\caption{Neutron matter equation of state in the Skyrme framework, SLy230b \citep{Chabanat}.}
\label{fig1}
\end{figure*}

\begin{figure*}
\includegraphics{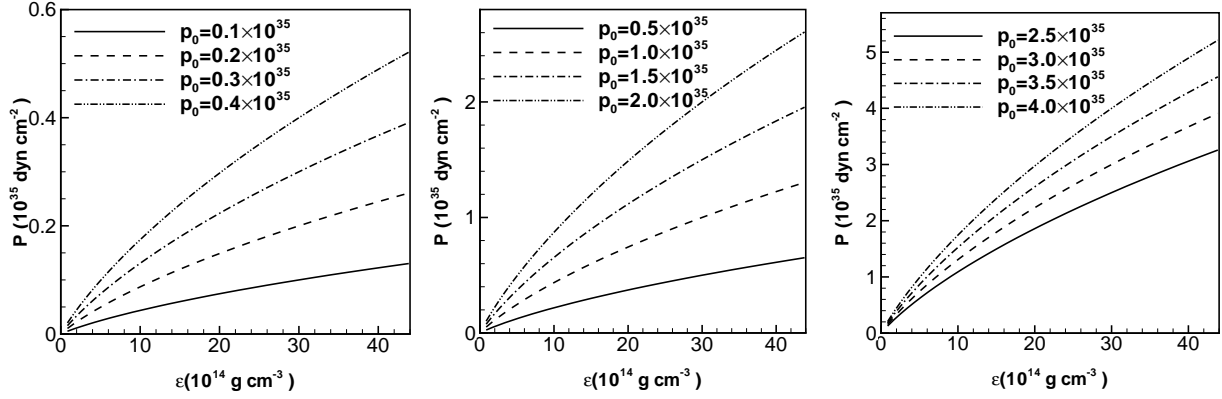}
\caption{Dark matter equation of state from the velocity profile of galaxies with the value ${\rho}_0=0.3\times10^{16}g\ cm^{-3}$ and different values of $p_0$ in $dyn\ cm^{-2}$ unit.}
\label{fig2}
\end{figure*}

In this study, we employ the NM EOS that has been derived
in the Skyrme framework \citep{Chabanat}. This equation is the result of the
Skyrme effective force, SLy230b, with the
improvement in the behavior with respect to the isospin degree
of freedom \citep{Chabanat}.
The EOS of NM in the SLy230b
parametrization is shown in Figure \ref{fig1}.
Recently, the DM EOS has been presented using the velocity profile of galaxies \citep{Barranco}.
In this study, we apply the pseudo isothermal model in
which the DM density profile determines the velocity profile \citep{Barranco}.
The EOS for the pseudo-isothermal
density profile is as follows \citep{Barranco},
\begin{eqnarray}\label{213}
       {p}({\rho})=\frac{8  {p}_0}{\pi^2-8}[\frac{\pi^2}{8}-\frac{arctan\sqrt{\frac
       {{\rho}_0}{{\rho}}-1}}{\sqrt{\frac{{\rho}_0}
       {{\rho}}-1}}
	-\frac{1}{2}(arctan\sqrt{\frac
       {{\rho}_0}{{\rho}}-1}\ )^2].
 \end{eqnarray}
In the above equation, ${\rho}$ and $p$ are the density and pressure of DM, respectively.
In addition, ${\rho}_0$ and ${p}_0$ are the only free parameters of the EOS, which for the DM in galaxies correspond to the
central density and pressure, respectively \citep{Barranco}. It has been found that this EOS has a functional dependence that is universal
for all galaxies, and the mentioned free parameters are related to the evolution
history of the galaxy. Using the EOS in Equation (\ref{213}) and the rotational curve data,
one can predict the central pressure and density of the galaxies. Here, we suppose that
the DM in DMANSs also behaves according to this EOS.
However, one should be careful about the value of the free parameters ${\rho}_0$ and ${p}_0$ in the EOS.
A logical choice that is similar to the case of galaxies is that the values of ${\rho}_0$ and ${p}_0$
are of the order of the central density and pressure of neutron stars, respectively, i.e., $\sim10^{15}g\ cm^{-3}$ for the density and $\sim10^{36}dyn\ cm^{-2}$ for the pressure.
We consider 12 EOSs with the value ${\rho}_0=0.3\times10^{16}g\ cm^{-3}$ and different values
of ${p}_0$, from $0.1\times10^{35}$ to $4.0\times10^{35}dyn\ cm^{-2}$. These EOSs are presented
in Figure \ref{fig2}. We should note that we have used units in which G = c = 1, and therefore the mass density and energy density of DM are equal, i.e., $\varepsilon=\rho$.

\begin{figure*}
\includegraphics{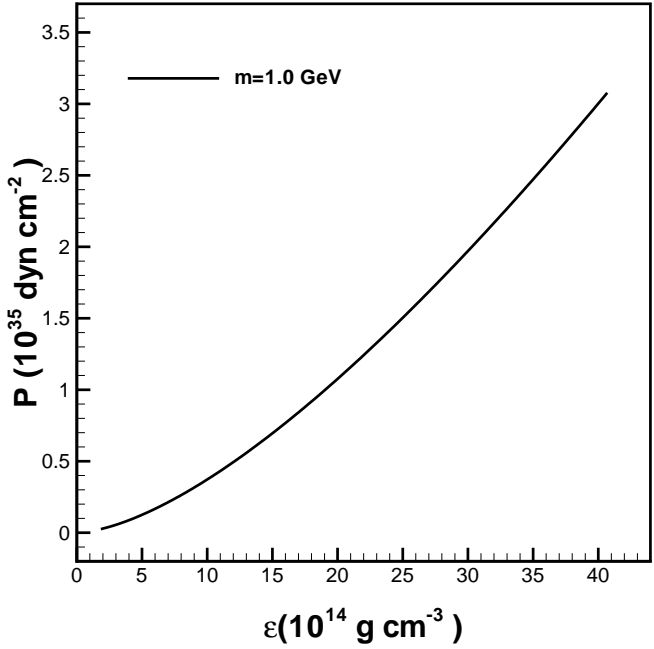}
\caption{Equation of state for a free gas of fermions at zero
temperature with the value of $m=1.0\ GeV$ for the mass of fermions.}
\label{fig12}
\end{figure*}

It is clear that
the increase in the value of ${p}_0$ leads to more stiffness in the EOS.
In addition, it is possible to consider DMANSs in which the DM sector is composed by free fermions with arbitrary masses \citep{Narain}. In this case the EOS at zero
temperature, $p(\rho)$, is obtained as follows \citep{Narain},
\begin{eqnarray}
\rho=\frac{1}{\pi^2}\int_0^{k_F} k^2 \sqrt{m^2+k^2} dk=\frac{m^4}{8 \pi^2}[(2z^3+z)(1+z^2)^{1/2}-sinh^{-1}(z)],
 \end{eqnarray}
\begin{eqnarray}
p=\frac{1}{3 \pi^2}\int_0^{k_F} \frac{k^4}{\sqrt{m^2+k^2}} dk=\frac{m^4}{24 \pi^2}[(2z^3-3z)(1+z^2)^{1/2}+3sinh^{-1}(z)].
 \end{eqnarray}
In the above equation, $k_F=(3\pi^2 n)^{1/3}$ denotes the Fermi momentum and n is the total number density of fermions. In addition, $m$ shows the mass of fermions and $z=k_F/m$. The EOS with mass $m=1.0\ GeV$ for the fermions is presented in Figure \ref{fig12}.

In this work, we study a DM admixed neutron star with two concentric spheres:
one containing the NM and the other the DM.
The EOS of NM
in the NM sphere is presented in Figure \ref{fig1}, and the EOSs of DM
in the DM sphere are given in Figures \ref{fig2} and \ref{fig12}.
In our calculations for the NM sphere, in the crust at densities
lower than $0.05\ fm^{-3}$, we use the EOS calculated by \cite{Baym}.
The contribution of DM in the DMANS can be different and depends on the history of the star \citep{Sandin} and the way that it originates \citep{Xiang}.
The gravitational field of the normal star in all phases of evolution can affect the DM and capture it into the star.
This is more probable when the star is more compact.
In contrast, the opposite scenario in which the DM objects trap the baryons has been also proposed \citep{Ciarcelluti}.
In this case, a more significant contribution of DM in the DMANS is expected. In this study, we explore DMANSs
with different proportions of DM in the center of the DMANS. This will be done by considering different central pressure ratios, i.e., central DM pressure to central NM pressure ratio
$\delta=\frac{p_D(r=0)}{p_N(r=0)}$, for the DMANS. The zero pressure ratio, $\delta=0$, shows a normal neutron star.


In our calculations of the DMANS structure, we apply the two-fluid formalism
for the NM and DM presented by \citep{Sandin,Ciarcelluti}.
In this formalism, it is proposed that DM interacts with NM just through gravity.
One static and spherically symmetric space-time with the line element, (we use the units in which G = c = 1),
\begin{eqnarray}
      d\tau^2=e^{2\nu(r)}dt^2-e^{2\lambda(r)}dr^2-r^2(d\theta^2+sin^2\theta d\phi^2),
 \end{eqnarray}
and the energy-momentum tensor of a perfect fluid,
\begin{eqnarray}
      T^{\mu \nu}=-p g^{\mu \nu}+(p+\varepsilon)u^{\mu}u^{\nu},
 \end{eqnarray}
are considered. In the above equation, $p$ and $\varepsilon$ are the total pressure and total energy density, respectively. The total energy density is related to the rest-mass energy density ($\varepsilon_{rm}$) and internal energy density ($\varepsilon_{in}$) by
 \begin{eqnarray}
      \varepsilon= \varepsilon_{rm}+\varepsilon_{in}.
 \end{eqnarray}
In our system, the total pressure and total energy density are the results of
both NM and DM quantities,
  \begin{eqnarray}
      p(r) = p_N(r) + p_D(r),
 \end{eqnarray}
   \begin{eqnarray}
      \varepsilon(r) =\varepsilon_N(r) + \varepsilon_D(r),
 \end{eqnarray}
in which $p_i$ and $\varepsilon_i$ are the pressure and energy density of neutron ($i=N$) and dark ($i=D$) matter at position $r$ from the center of the star, respectively.
By applying some calculations, the Einstein field equations lead to \citep{Sandin,Ciarcelluti,Xiang},
\begin{eqnarray}
      e^{-2\lambda(r)}=1-\frac{2M(r)}{r},
       \end{eqnarray}
\begin{eqnarray}
      \frac{d\nu}{dr}=\frac{M(r)+4\pi r^3 p(r)}{r[r-2M(r)]},
       \end{eqnarray}
\begin{eqnarray}
            \frac{dp_N}{dr}=-[p_N(r)+\varepsilon_N(r)] \frac{d\nu}{dr},
 \end{eqnarray}
\begin{eqnarray}
            \frac{dp_D}{dr}=-[p_D(r)+\varepsilon_D(r)] \frac{d\nu}{dr}.
 \end{eqnarray}
In the above equations, $r$ denotes the radial coordinate from the center of the star. $M(r)=\int_0^r dr 4 \pi r^2 \varepsilon(r)$ is the total mass inside a sphere with radius $r$.
This separation for DM and NM is due to the assumption that the two sectors interact just via gravity. The structure of DMANS is described by these two-fluid TOV equations.
The condition for the DMANS radius, $R$, and
mass, $M(R)$, is $p(R) = 0$ \citep{Xiang}. Moreover, the radius and mass
of the NM sphere and DM sphere are determined with
the conditions $p_N(R_{N})=0$ and $p_D(R_{D})=0$, respectively \citep{Xiang}.
It should be noted that the radial dependencies of the
pressure and density are different in the two sectors.

\section{Results and discussion} \label{sec:floats}

\begin{figure*}
\includegraphics{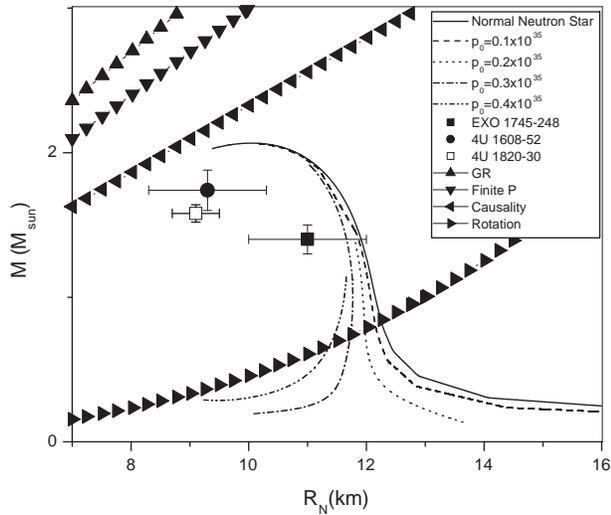}
\caption{ $M-R_{N}$ relation for a normal neutron star and DMANSs with different values of ${p}_0$ in units of $dyn\ cm^{-2}$ for the case of $\delta=1$ and some observational data.
Curves that exclude some regions, i.e., by the general relativity $M>\frac{c^2 R}{2 G}$ (GR), by the finite pressure $M>\frac{4}{9}\frac{c^2 R}{G}$ (Finite P), by causality $M>\frac{10}{29}\frac{c^2 R}{G}$ (Causality), and by the rotation of 716 Hz pulsar J1748-2446ad (Rotation), are also presented \citep{Lattimer}.
}
\label{fig3}
\end{figure*}

\begin{figure*}
\includegraphics{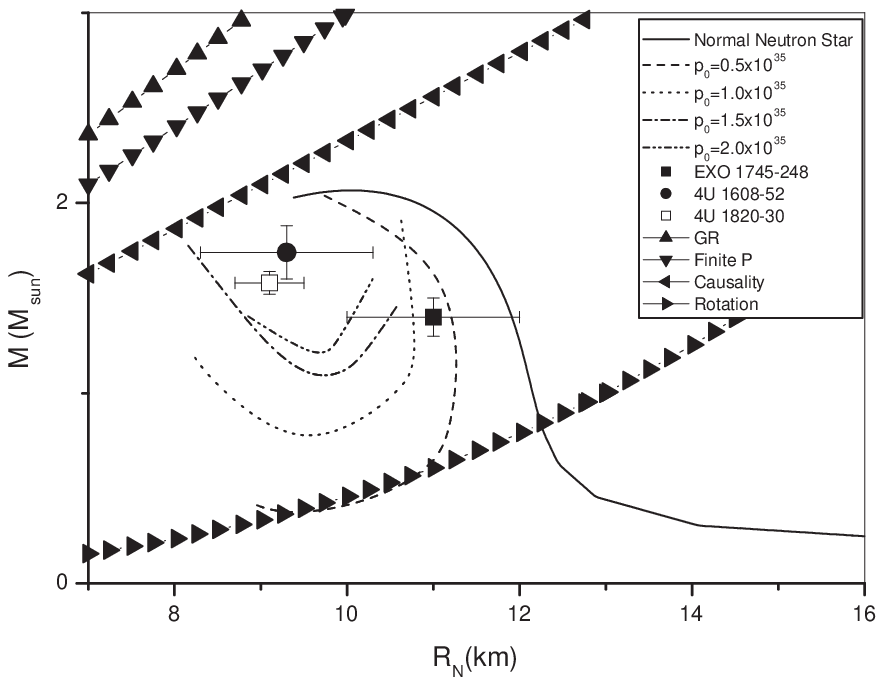}
\caption{Same as Figure \ref{fig3} but for other values of ${p}_0$ in $dyn\ cm^{-2}$ unit.}
\label{fig4}
\end{figure*}

\begin{figure*}
\includegraphics{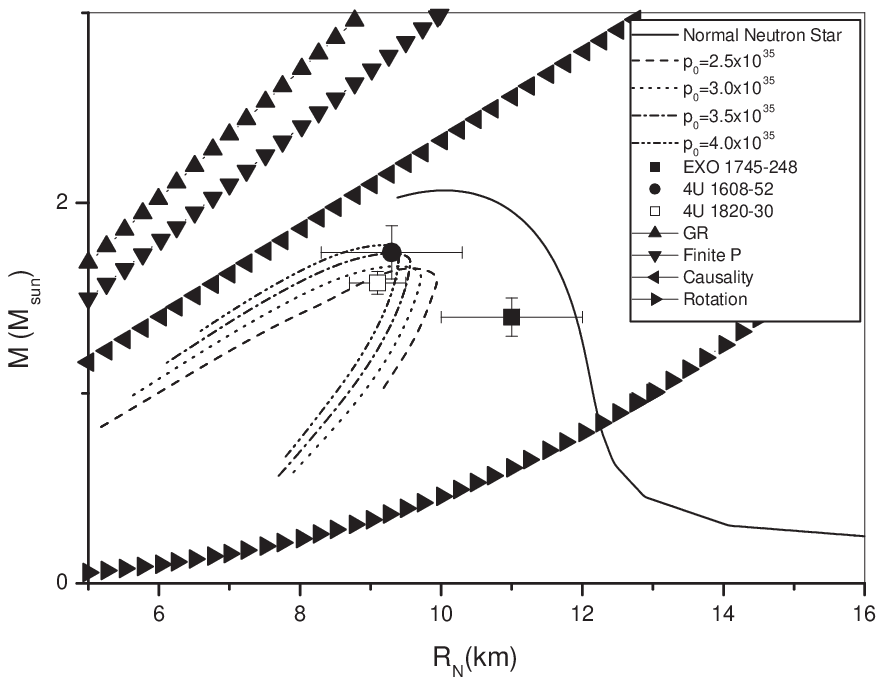}
\caption{Same as Figure \ref{fig3} but for other values of ${p}_0$ in $dyn\ cm^{-2}$ unit.}
\label{fig5}
\end{figure*}

\begin{figure*}
\includegraphics{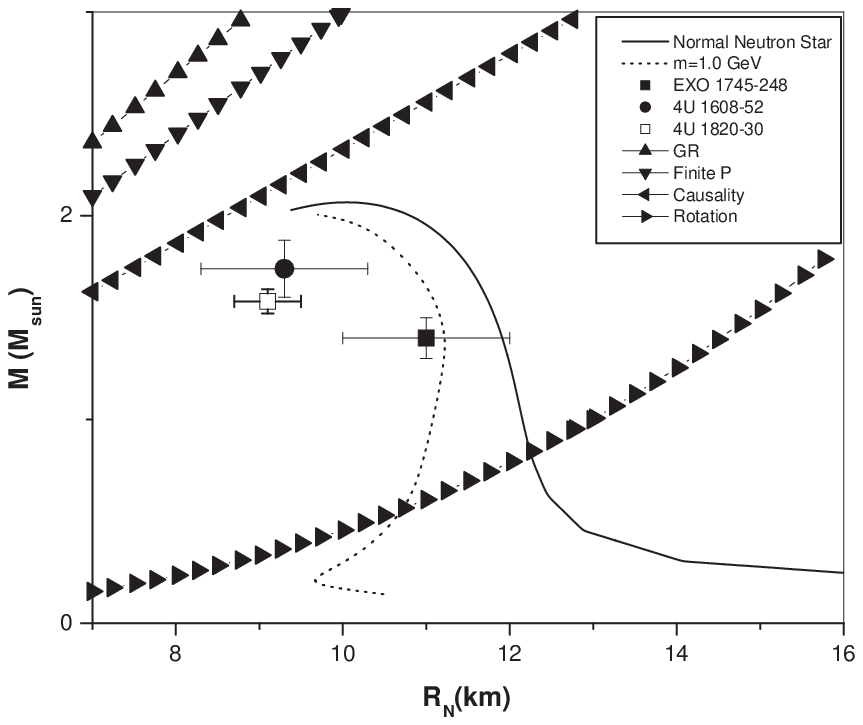}
\caption{Same as Figure \ref{fig3} but for the DMANS with the DMEOS given in
Figure \ref{fig12}.}
\label{fig31}
\end{figure*}

Figures \ref{fig3}-\ref{fig31} show the total mass of the DMANS versus the radius of the NM sphere ($M-R_{N}$ relation) for a normal neutron star and DMANSs with different DM equations of state (DMEOSs) when the central pressures of NM and DM are the same, i.e., $\delta=1$. The curves that apply constraints for the mass and radius, i.e., by the general relativity $M>\frac{c^2 R}{2 G}$ (GR), by the finite pressure $M>\frac{4}{9}\frac{c^2 R}{G}$ (Finite P), by causality $M>\frac{10}{29}\frac{c^2 R}{G}$ (Causality), and by the rotation of 716 Hz pulsar J1748-2446ad (Rotation), are also presented  \citep{Lattimer}. The mass-radius relations that pass in the permitted region are acceptable.
We can see that with all DMEOSs
 the radius of a DMANS
is smaller than the radius of a normal neutron
star with the same mass. This decrease of radius in DMANSs
is in agreement with previous results \citep{Leung11,Leung12,Xiang}.
It is obvious that the $M-R_{N}$ relations obtained using DMEOSs
corresponding to ${p}_0 \leq 0.2\times10^{35}dyn\ cm^{-2}$ show a behavior like that of normal neutron stars that are gravitationally bound.
However, the other DMEOSs, which are the stiffer equations of state, lead to $M-R_{N}$ relations that are significantly different from the normal neutron
stars. Such $M-R_{N}$ relations with different slopes, representing the self-bound DMANSs,
are the result of the interaction between DM and NM.
The significant effects of the stiff DM equations of state on $M-R_{N}$ relations have been also mentioned in \cite{Xiang}.
Our results indicate that the $M-R_{N}$ relation for the DMEOS with
 ${p}_0 = 0.5\times10^{35}dyn\ cm^{-2}$ is approximately in agreement with the result related to the DMEOS with $m=1.0\ GeV$.
Figures \ref{fig3}-\ref{fig31} show that, for some $M-R_{N}$ relations, with a value of $R_{N}$ two possible
DMANSs with different masses can exist. This is a result of different degrees
of DM domination in DMANSs. Put differently, two equal sized DMANSs that are affected by the DM differently have different masses.
For example, consider a $M-R_{N}$ relation with ${p}_0 = 0.5\times10^{35}dyn\ cm^{-2}$.
With this DMEOS, two DMANSs with similar radii $11.25$ and $11.36\ km$ have masses of about $0.77M_{\odot}$ and $1.67M_{\odot}$, respectively.
The ratios of the mass of the DM sphere to the total mass for these two DMANSs
are $\frac{M_D}{M}=0.13$ and $\frac{M_D}{M}=0.02$ respectively, indicating the greater effect of DM in the former.
In addition, in the case of $m=1.0\ GeV$, two DMANSs with radii of $10.41$
and $10.39\ km$ have masses of $0.41M_{\odot}$ and $1.95M_{\odot}$, and the ratios of $\frac{M_D}{M}$ are 0.29 and 0.03, respectively.
\begin{table*}
\caption{The Maximum Masses and Corresponding Visible Radii for A Normal Neutron
star and DMANSs with
Different Dark Matter Equations of State for the Case of $\delta=1$}
\label{table1}       
\begin{center}  {\footnotesize
\begin{tabular}{|c|c|c|}
\hline& $M_{max} (M_{\odot})$  & \multicolumn{1}{c|}{R (km)}  \\\hline
Normal Neutron Star &2.07& 10.04   \\\hline
 ${p}_0(10^{35}dyn\ cm^{-2})$ & &    \\
 $0.1$ &2.06 &10.18   \\
 $0.2$& 1.39& 11.80 \\
 $0.3$ &1.92 &11.02   \\
 $0.4$ & 1.16& 11.66 \\
 $0.5$ &2.05  &9.71    \\
 $1.0$ &1.91  & 10.63  \\
 $1.5$ &1.78  & 8.16   \\
 $2.0$ & 1.60 & 10.30  \\
 $2.5$ &   1.68& 9.35  \\
 $3.0$&1.68 & 9.38  \\
 $3.5$ &  1.74 &  9.25   \\
 $4.0$ &  1.78 &  9.11  \\\hline
 m (GeV) & &    \\
  $1.0$ & 2.01  & 9.70   \\
 \hline
\end{tabular} }
\end{center}
\end{table*}
The maximum mass of DMANS depends on the DMEOSs (see Table~\ref{table1}).
We can see that with all DMEOSs, the maximum mass of the DMANS is lower than that of the normal neutron star. The decrease of the maximum mass due to the DM has been also reported in other studies \citep{Leung11,Leung12,Li,Xiang}.
Figures \ref{fig3}-\ref{fig31} also show the comparison of observational data with the theoretical results.
It is obvious that EXO 1745-248, with mass
$M = 1.4 \pm 0.1\ M_{\odot}$ and radius $R = 11\pm 1\ km$, can hardly be modeled with the normal neutron star. However, our model for DMANSs with $0.5\times10^{35}dyn\ cm^{-2}\leq {p}_0 \leq1.5 \times10^{35}dyn\ cm^{-2}$ or with $m=1.0\ GeV$
can explain this observational result.
The neutron star in 4U 1820-30 with $M = 1.58\pm0.06\ M_{\odot}$ and $R = 9.1 \pm 0.4\ km$ and the neutron star
in 4U 1608-52 with $M = 1.74\pm0.14\ M_{\odot}$  and $R =  9.3 \pm 1.0\ km$ cannot be normal neutron stars.
It is clear that the DMEOSs with $2.5\times10^{35}dyn\ cm^{-2}\leq {p}_0\leq4.0 \times10^{35}dyn\ cm^{-2}$ corresponding to self-bound stars lead to results that are in agreement with the properties of the neutron stars
in 4U 1608-52
and 4U 1820-30.
\begin{table*}
\caption{The Mass and Radius of the Dark and Neutron Matter Spheres of the Maximum
Mass DMANSs with Different DMEOSs for the Central Pressure Ratio $\delta=1$}
\label{table6}       
\begin{center}  {\footnotesize
\begin{tabular}{|c|c|c|c|c|c|}
\hline ${p}_0(10^{35}dyn\ cm^{-2})$  &$M_{D} (M_{\odot})$  & \multicolumn{1}{c|}{$R_{D} (km)$} &
\multicolumn{1}{c|}{$M_{N} (M_{\odot})$}&\multicolumn{1}{c|}{$R_{N} (km) $}&$M_{max} (M_{\odot})$  \\\hline
$0.1$&$<0.01$ & 1.85& 2.06& 10.18&2.06 \\
 $0.2$&0.02&4.34&1.37 &11.80& 1.39 \\
 $0.3$&0.01&3.48&1.90 & 11.02&1.92   \\
 $0.4$&0.05&5.86 &1.11&11.66& 1.16\\
 $0.5$&0.01 &2.70& 2.04& 9.71&2.05   \\
 $1.0$&0.06&5.16&1.85& 10.63&1.91  \\
 $1.5$&1.63&19.95&0.15&8.16&1.78  \\
 $2.0$&0.24&8.35&1.36&10.30& 1.60  \\
 $2.5$&0.99&15.92&0.69&9.35&   1.68 \\
 $3.0$&0.34&9.38&1.34&9.38 &1.68   \\
 $3.5$&0.34&9.25&1.40&9.25  &  1.74   \\
 $4.0$&0.34&9.11&1.44& 9.11 &  1.78  \\\hline
   $m\ (GeV)$& & & & &   \\
 $1.0$&0.05&3.26&1.96&9.70 &2.01  \\
\hline
\end{tabular} }
\end{center}
\end{table*}

Table~\ref{table6} presents the mass and radius of the dark and NM spheres
relating to the DMANSs with the maximum
mass. It can be seen that
for DMANSs with DMEOSs corresponding to ${p}_0 \leq 1.0 \times
10^{35}dyn\ cm^{-2}$, the significant contribution in the mass
is related to the NM sphere. In addition, in these cases,
the radius of the DM sphere is smaller than
the NM sphere.
Therefore, for ${p}_0 \leq 1.0 \times
10^{35}dyn\ cm^{-2}$, the star is
NM-dominated and gravitationally bound.
However, in the DMANS with ${p}_0 \simeq 1.5 \times 10^{35}dyn\ cm^{-2}$,
the DM sphere with the mass $1.63\ M_{\odot}$ is heavier
than the NM sphere with the mass $0.15\ M_{\odot}$.
In this star, the DM halo of radius $19.95\ km$
surrounds the NM sphere of smaller radius $8.16\ km$. This is
a DM-dominated neutron star. We can conclude from the above discussion
that the neutron star EXO 1745-248 could be a DM-dominated neutron star.
Figures \ref{fig4} and \ref{fig31} indicate that both DMEOSs ${p}_0 \simeq 1.0 \times 10^{35}dyn\ cm^{-2}$ and
$m=1.0\ GeV$, which lead to self-bound DMANSs, can be used to explain the observational result of EXO 1745-248. Moreover, Table~\ref{table6} shows that
the masses of a DM sphere obtained from these two models, i.e., $0.06\ M_{\odot}$
and $0.05\ M_{\odot}$ respectively, are approximately the same.
In addition, the DMEOS with ${p}_0 \simeq 2.5 \times 10^{35}dyn\ cm^{-2}$
leads to a DMANS with a heavier and larger
DM sphere compared to the NM sphere.
Consequently, this star is also
a DM-dominated neutron star. Therefore, the neutron stars 4U 1608-52
and 4U 1820-30 could be self-bound and DM-dominated neutron stars.
Table~\ref{table6} indicates that the DMANSs with DMEOSs corresponding to
${p}_0 \geq 3.0\times 10^{35}dyn\ cm^{-2}$ are
stars with a more massive NM sphere, but have two spheres of equal radius.
By increasing the value of the free parameter ${p}_0$ from
$3.0$ to $4.0$, the mass of the NM sphere increases while the mass of the DM
sphere remains constant at $0.34\ M_{\odot}$. The increase in the value of ${p}_0$ reduces the radius of the
NM and DM spheres, leading to the compactness of the star.

\begin{figure*}
\includegraphics{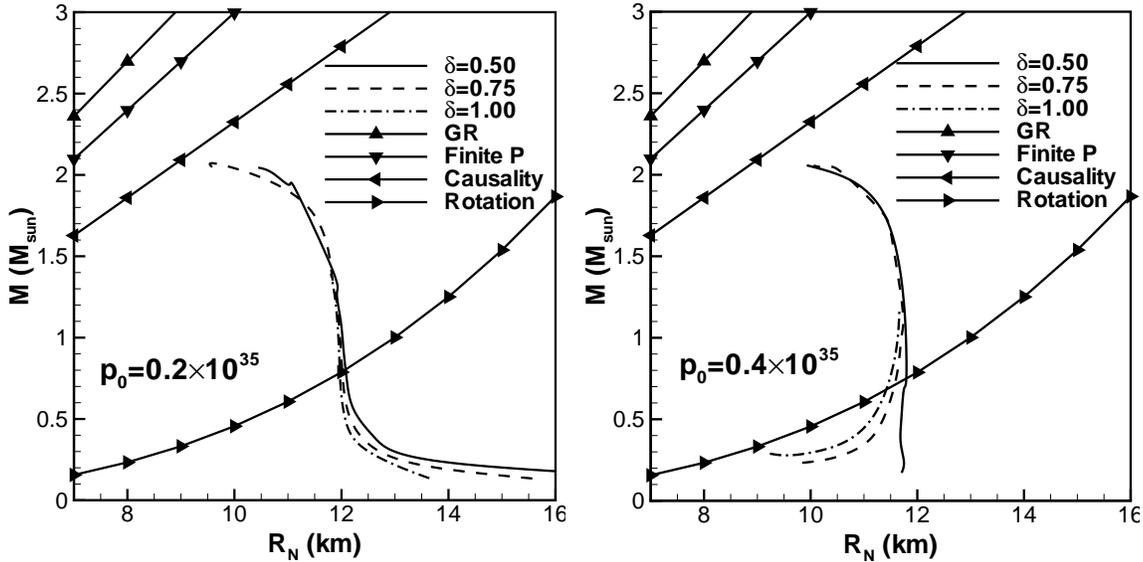}
\caption{ $M-R_{N}$ relations and constraint curves
with two values of ${p}_0$ in units of $dyn\ cm^{-2}$ for different central pressure ratios, $\delta$.}
\label{fig6}
\end{figure*}
\begin{figure*}
\includegraphics{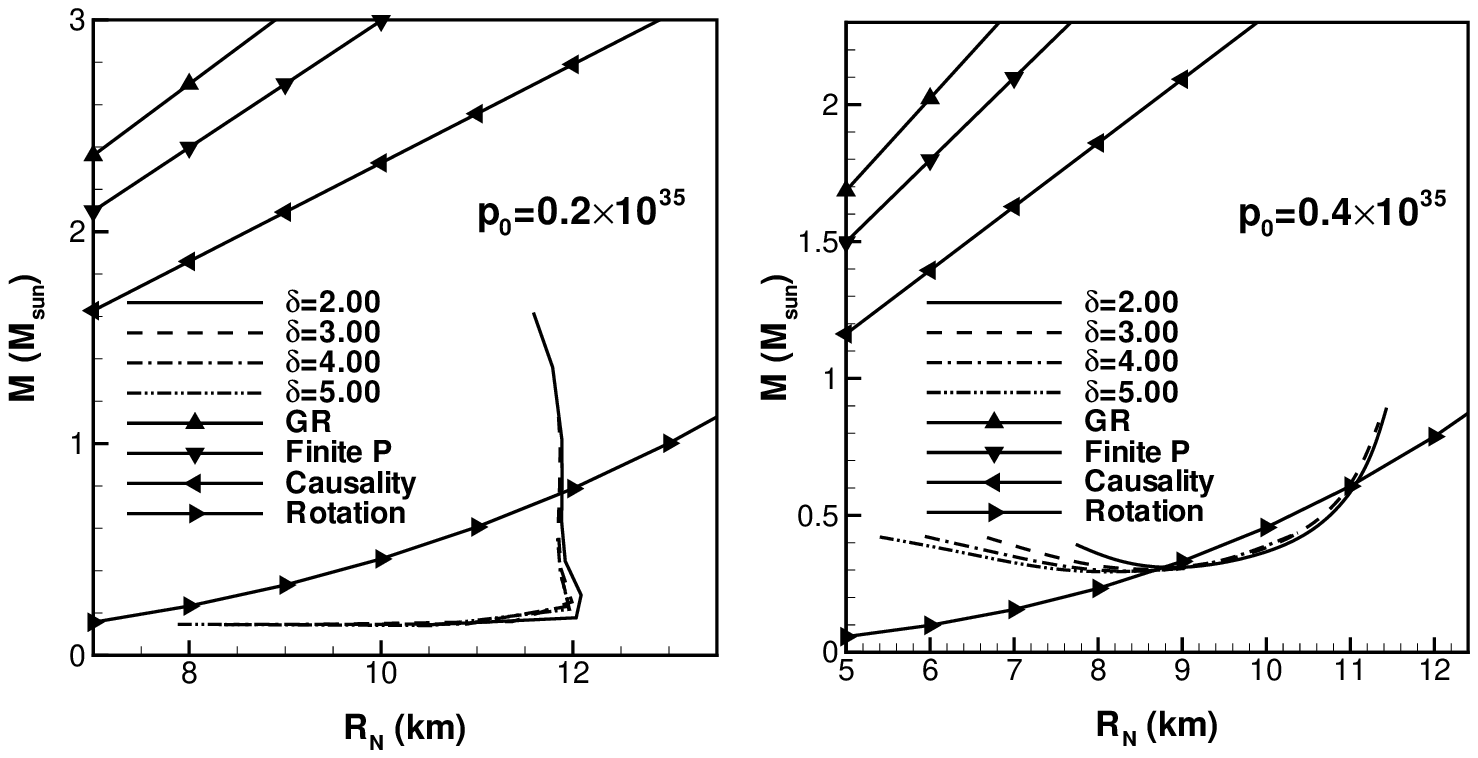}
\caption{Same as Figure \ref{fig6} but for other values of central pressure ratio, $\delta$.}
\label{fig7}
\end{figure*}
\begin{figure*}
\includegraphics{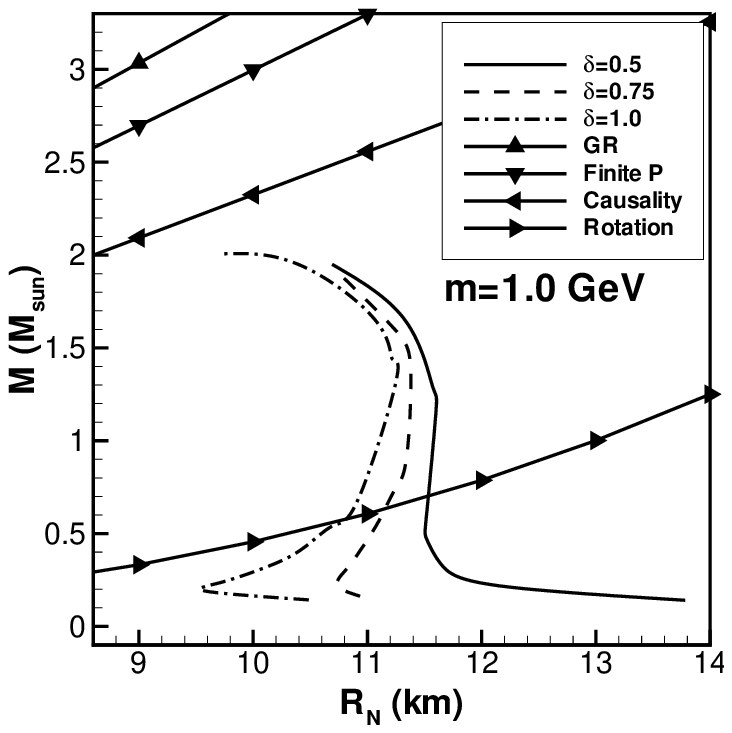}
\includegraphics{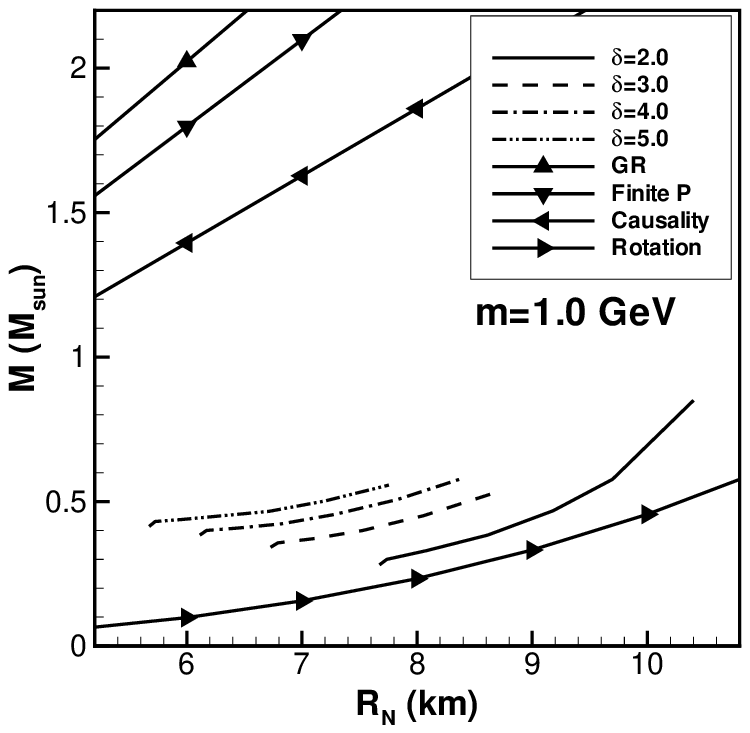}
\caption{ $M-R_{N}$ relations and constraint curves
with the fermionic DMEOS for different central pressure ratios, $\delta$.}
\label{fig61}
\end{figure*}

Figures \ref{fig6}-\ref{fig61} show the $M-R_{N}$ relations
with different DMEOSs for various central pressure ratios, $\delta$.
It is obvious that, for all DMEOSs, when the central pressure ratio grows from $0.5$ to $1$, the low mass DMANSs (about $M \leq M_{\odot}$) will become more compact.
This effect is more significant for the case of $m=1.0\ GeV$.
In addition, by increasing the central pressure ratio from $2$ to $5$,
for DMANSs with ${p}_0 =0.2 \times10^{35}dyn\ cm^{-2}$ the radius of
the star is not affected. However, low mass DMANSs with ${p}_0 =0.4 \times10^{35}dyn\ cm^{-2}$ and $m=1.0\ GeV$ become more compact. DMANSs with $m=1.0\ GeV$ are more affected by the value of central pressure ratio compared to the cases with ${p}_0 =0.4 \times10^{35}dyn\ cm^{-2}$.
In the case with $m=1.0\ GeV$, DMANSs with small central pressure ratios, i.e., $\delta \leq 0.75$, are like normal neutron stars and gravitationally bound. However, for higher values of central pressure ratio the DMANSs  are self-bound stars.
We have found from Figure \ref{fig7} that DMANSs with the DMEOS
${p}_0=0.2\times10^{35}dyn\ cm^{-2}$ and central pressure ratios
$\delta=4.00$ and $\delta=5.00$ do not exist.
Figures \ref{fig6}-\ref{fig61} confirm that
the $M-R_{N}$ relation can be influenced significantly by the central pressure ratio, similar to the results of \citep{Xiang}.
\begin{table*}
\caption{The Maximum Masses and Corresponding Visible Radii for DMANSs with DMEOS Corresponding to ${p}_0=0.2\times10^{35}dyn\ cm^{-2}$ and Different Values of Central Pressure Ratio, $\delta$ }
\label{table2}       
\begin{center}  {\footnotesize
\begin{tabular}{|c|c|c|}
\hline  $\delta$ &$M_{max} (M_{\odot})$  & \multicolumn{1}{c|}{$R (km)$}  \\\hline
0.50 &2.05& 10.44   \\
0.75&2.06 &9.81  \\
1.00& 1.39& 11.80 \\
2.00 &1.62 &11.59  \\
3.00 & 1.18& 11.84 \\
\hline
\end{tabular} }
\end{center}
Note. In this case, DMANSs with $\delta\geq4.00$ do not exist.

\vspace*{2cm}  
\end{table*}
\begin{table*}
\caption{Same as Table~\ref{table2} but for DMANSs with DMEOS Corresponding to ${p}_0=0.4\times10^{35}dyn\ cm^{-2}$}
\label{table3}       
\begin{center}  {\footnotesize
\begin{tabular}{|c|c|c|}
\hline  $\delta$ &$M_{max} (M_{\odot})$  & \multicolumn{1}{c|}{$R (km)$}  \\\hline
0.50 &2.06& 9.93    \\
0.75&2.06 &10.05  \\
1.00& 1.16& 11.66 \\
2.00 &0.89 &11.43   \\
3.00 &0.84 & 11.34 \\
4.00 &0.42 &5.94  \\
5.00 & 0.42& 5.40 \\
\hline
\end{tabular} }
\end{center}
\end{table*}
\begin{table*}
\caption{Same as Table~\ref{table2} but for DMANSs with Fermionic DMEOS Corresponding to $m=1.0\ GeV$}
\label{table31}       
\begin{center}  {\footnotesize
\begin{tabular}{|c|c|c|}
\hline  $\delta$ &$M_{max} (M_{\odot})$  & \multicolumn{1}{c|}{$R (km)$}  \\\hline
0.50 &1.95 & 10.69   \\
0.75& 1.88 & 10.79 \\
1.00&2.01  &9.70   \\
2.00 &0.85  &10.40    \\
3.00 &0.53  & 8.65  \\
4.00 &0.58  & 8.41  \\
5.00 & 0.56 &7.75   \\
\hline
\end{tabular} }
\end{center}
\end{table*}
The results for the maximum masses and corresponding radii are presented in Tables~\ref{table2}-\ref{table31}. It can be seen from Tables~\ref{table2} and \ref{table3} that, for DMANSs with ${p}_0 =0.2 \times10^{35}dyn\ cm^{-2}$ and $\delta\geq2$ as well as the DMANSs with ${p}_0 =0.4 \times10^{35}dyn\ cm^{-2}$, the maximum mass decreases by increasing the central pressure ratio, in agreement with the results of \citep{Xiang}. Table~\ref{table31} also indicates that, in the case of fermionic DMEOS, $m=1.0\ GeV$, the maximum masses are smaller for $\delta>1$ than the cases with $\delta<1$.
\begin{figure*}
\includegraphics{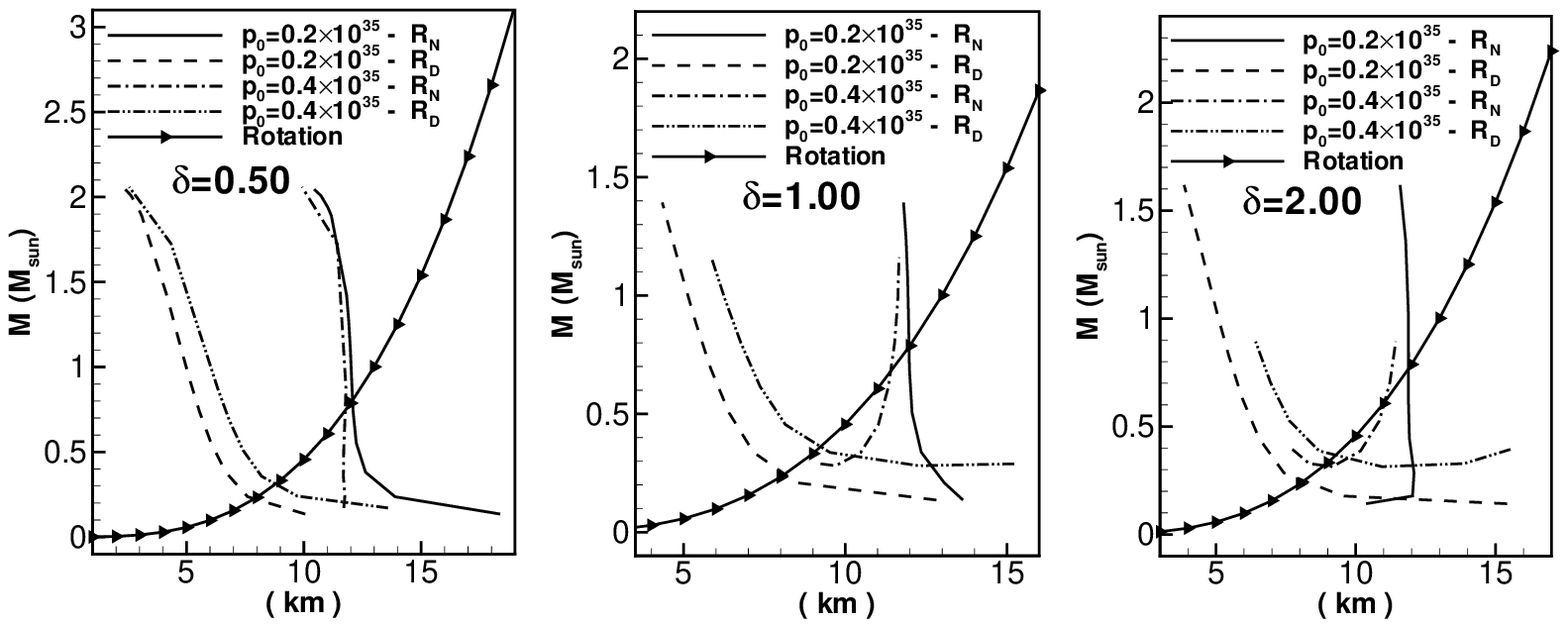}
\caption{The total mass vs. neutron ($R_{N}$) and dark ($R_{D}$) matter sphere radii of DMANSs, and the curve that excludes a region by the rotation
of 716 Hz pulsar J1748-2446ad (Rotation), with different values of ${p}_0$ in units of $dyn\ cm^{-2}$ and central pressure ratio, $\delta$.}
\label{fig8}
\end{figure*}
\begin{figure*}
\includegraphics{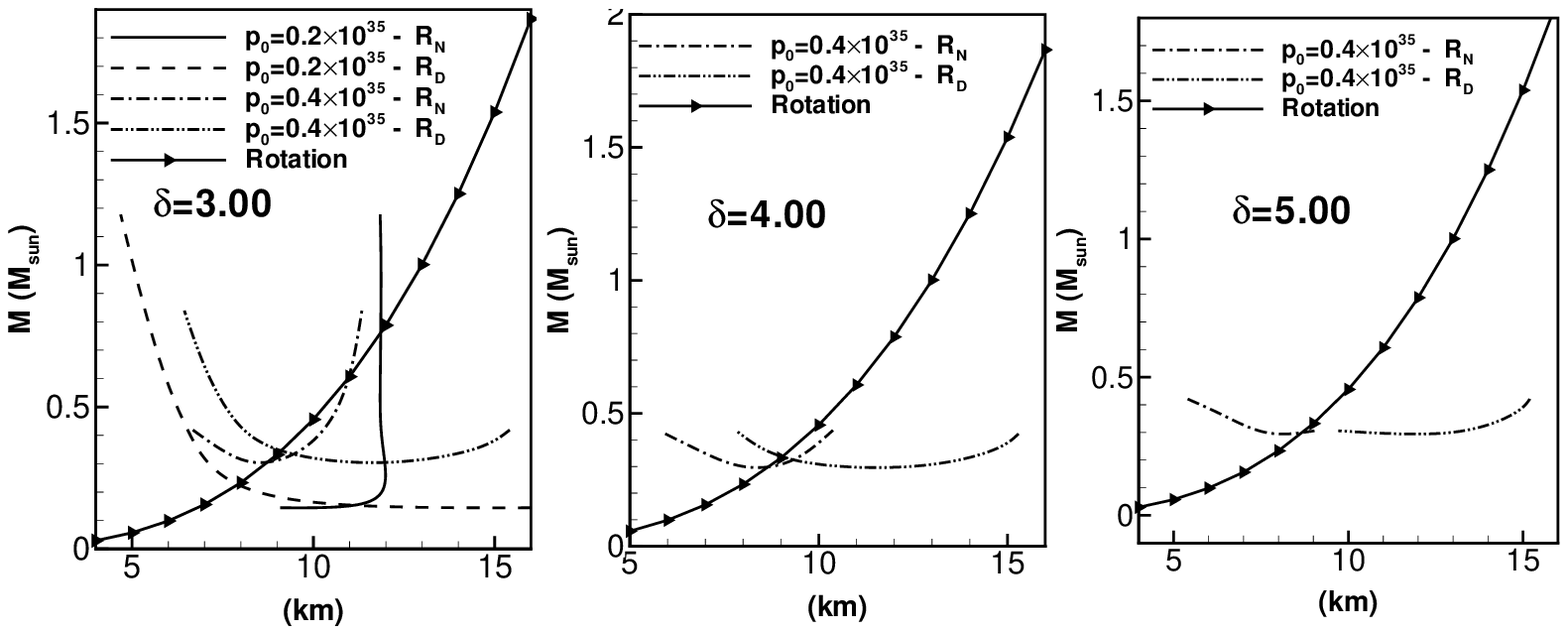}
\caption{Same as Figure \ref{fig8} but for other values of central pressure ratio, $\delta$.}
\label{fig9}
\end{figure*}
\begin{figure*}
\includegraphics{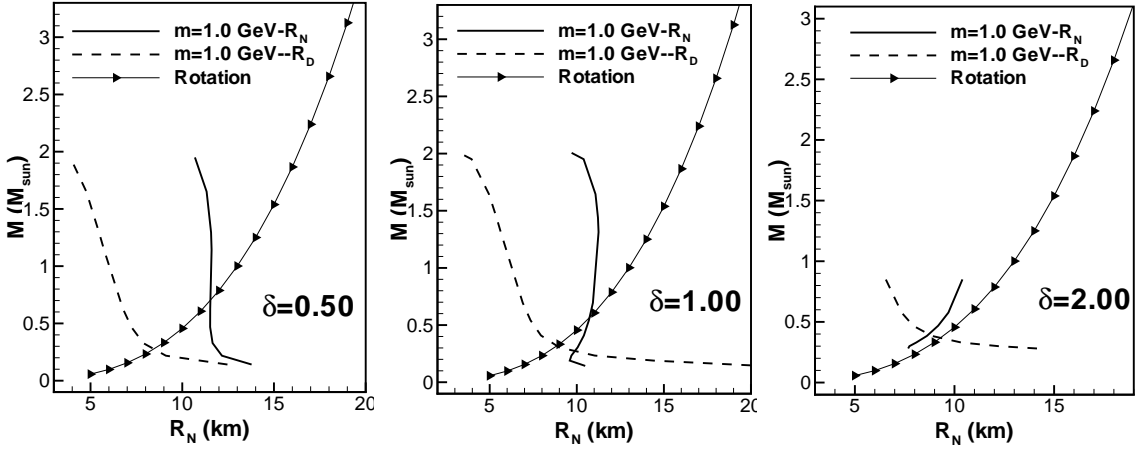}
\caption{The total mass vs. neutron ($R_{N}$) and dark ($R_{D}$) matter sphere radii of DMANSs, and the curve that excludes a region by the rotation with the fermionic DMEOS for different central pressure ratios, $\delta$.}
\label{fig81}
\end{figure*}
\begin{figure*}
\includegraphics{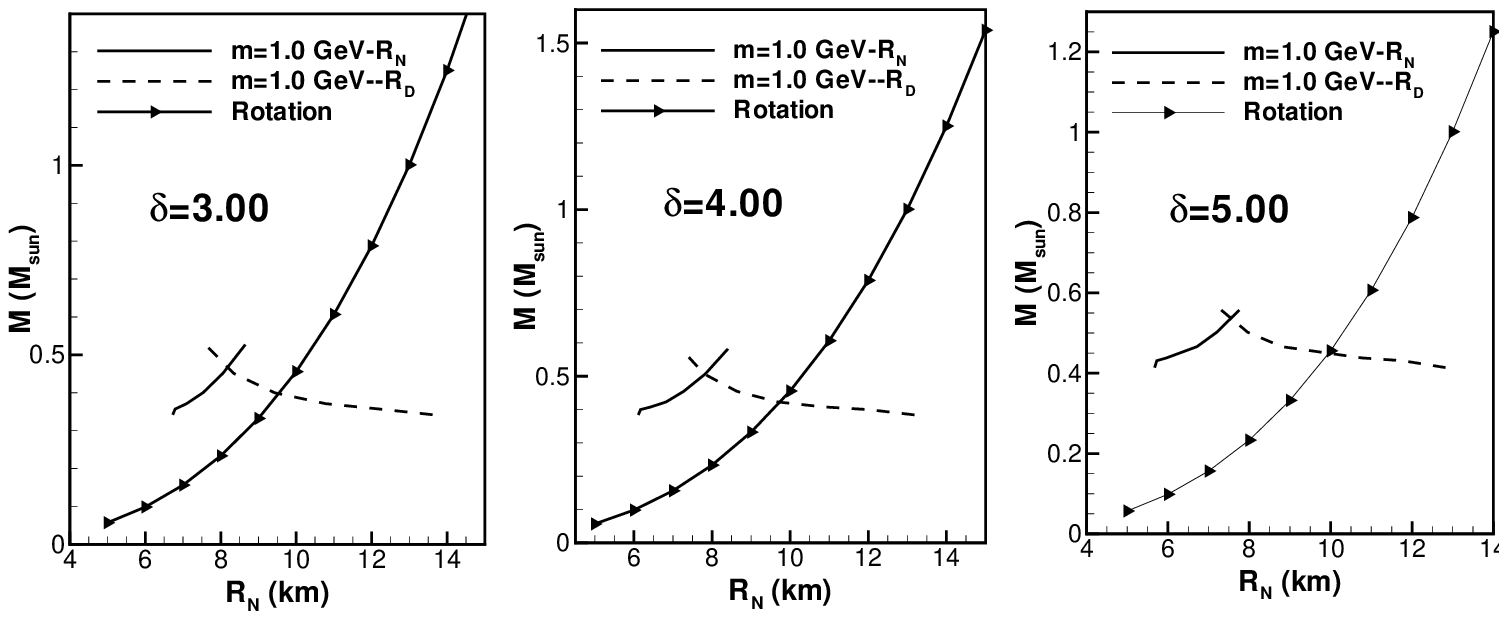}
\caption{Same as Figure \ref{fig81} but for other values of central pressure ratio, $\delta$.}
\label{fig91}
\end{figure*}

The total mass versus radius of the NM and DM spheres for different DMEOSs
and various values of central pressure ratio is given in Figures \ref{fig8}-\ref{fig91}.
These figures show the size of the
NM and DM spheres for a DMANS with a specific mass.
We can see that, for all values of central pressure ratio, the radius
of the NM sphere of the star with ${p}_0 =0.4 \times10^{35}dyn\ cm^{-2}$ is smaller than the case of ${p}_0 =0.2 \times10^{35}dyn\ cm^{-2}$
with the same mass.
However, it is clear that the radius of the DM sphere of the star with
${p}_0 =0.4 \times10^{35}dyn\ cm^{-2}$ is larger compared to the case with ${p}_0 =0.2 \times10^{35}dyn\ cm^{-2}$.
Figures \ref{fig8}-\ref{fig91} indicate that in the cases with ${p}_0 =0.2 \times10^{35}dyn\ cm^{-2}$ and ${p}_0 =0.4 \times10^{35}dyn\ cm^{-2}$ for $\delta \leq 3.0$, and
in the case of $m=1.0\ GeV$ for $\delta \leq 1.0$, the radius of the NM sphere is larger than the radius of the DM one.
The difference between the radius of the NM and DM spheres is larger
in stars with ${p}_0 =0.2 \times10^{35}dyn\ cm^{-2}$ compared to the case of ${p}_0 =0.4 \times10^{35}dyn\ cm^{-2}$.
For all DMEOSs, by increasing the central pressure ratio,
the radius of the NM sphere decreases while the radius of the DM sphere
increases. For stars with ${p}_0 =0.4 \times10^{35}dyn\ cm^{-2}$ and $\delta\geq3.00$, as well as stars with $m=1.0\ GeV$ and $\delta \geq 2.00$, we can see that there are low mass DMANSs in which the NM sphere is surrounded by a halo of DM, which is consistent with previous results \citep{Leung11,Leung12}.
It is evident that with the higher values of $\delta$ (for example in stars with ${p}_0 =0.4 \times10^{35}dyn\ cm^{-2}$ for $\delta\geq5$),
the NM sphere is surrounded by a halo of DM for all stars.
\begin{table*}
\caption{The Mass and Radius of the Dark and Neutron Matter Spheres of the Maximum
Mass DMANSs with DMEOS Corresponding to ${p}_0=0.2\times10^{35}dyn\ cm^{-2}$ for Different Values of Central Pressure Ratio, $\delta$}
\label{table4}       
\begin{center}  {\footnotesize
\begin{tabular}{|c|c|c|c|c|c|}
\hline  $\delta$ &$M_{D} (M_{\odot})$  & \multicolumn{1}{c|}{$R_{D} (km)$} &
\multicolumn{1}{c|}{$M_{N} (M_{\odot})$}&\multicolumn{1}{c|}{$R_{N} (km) $}&$M_{max} (M_{\odot})$  \\\hline
0.50 &0.01&2.39&2.04 &10.44  &2.05   \\
0.75&0.01 &1.96&2.05 &9.811&2.06   \\
1.00& 0.02& 4.34&1.37& 11.80 &1.39\\
2.00 &0.02 &3.86 &1.60 &11.59 &1.62  \\
3.00 &0.03&4.69&1.15&11.84  & 1.18\\
\hline
\end{tabular} }
\end{center}
\end{table*}
\begin{table*}
\caption{Same as Table~\ref{table4} but for DMANSs with DMEOS Corresponding to ${p}_0=0.4\times10^{35}dyn\ cm^{-2}$}
\label{table5}       
\begin{center}  {\footnotesize
\begin{tabular}{|c|c|c|c|c|c|}
\hline  $\delta$ &$M_{D} (M_{\odot})$  & \multicolumn{1}{c|}{$R_{D} (km)$} &
\multicolumn{1}{c|}{$M_{N} (M_{\odot})$}&\multicolumn{1}{c|}{$R_{N} (km) $} &$M_{max} (M_{\odot})$ \\\hline
0.50 &0.01& 2.56&2.05 &9.93  &2.06  \\
0.75&0.01 &2.75&2.05&10.05 &2.06  \\
1.00&0.05& 5.86&1.11&11.66& 1.16  \\
2.00 &0.08&6.42 &0.81 &11.43  &0.89  \\
3.00 &0.08 & 6.44&0.76&11.34&0.84 \\
4.00 &0.41&15.29&0.01&5.94&0.42 \\
5.00 & 0.41&15.19&0.01&5.40  & 0.42\\
\hline
\end{tabular} }
\end{center}
\vspace*{2cm}  
\end{table*}
\begin{table*}
\caption{Same as Table~\ref{table4} but for DMANSs with fermionic DMEOS corresponding to $m=1.0\ GeV$.}
\label{table41}       
\begin{center}  {\footnotesize
\begin{tabular}{|c|c|c|c|c|c|}
\hline  $\delta$ &$M_{D} (M_{\odot})$  & \multicolumn{1}{c|}{$R_{D} (km)$} &
\multicolumn{1}{c|}{$M_{N} (M_{\odot})$}&\multicolumn{1}{c|}{$R_{N} (km) $}&$M_{max} (M_{\odot})$  \\\hline
0.50 &0.04&3.90&1.91&10.69& 1.95   \\
0.75&0.06&4.27&1.82& 10.79&1.88  \\
1.00&0.05 &3.26&1.96& 9.70&2.01\\
2.00&0.18&6.55&0.67& 10.40& 0.85 \\
3.00 &0.25&7.60&0.28&8.65&0.53\\
4.00 &0.28&7.21&0.30&8.41&0.58 \\
5.00 &0.31&7.31&0.25& 7.75&0.56\\
\hline
\end{tabular} }
\end{center}
\end{table*}

Tables~\ref{table4}-\ref{table41} present the mass and radius of the dark and NM spheres
of the maximum mass DMANSs with different DMEOSs and for various values of central pressure ratio.
We can see that the mass of a DM sphere increases when
the central pressure ratio increases. This effect is more significant in the
case of ${p}_0 =0.4 \times10^{35}dyn\ cm^{-2}$. In addition, with this DMEOS the mass of the NM sphere
decreases as the central pressure ratio grows.
In the case with ${p}_0 =0.2 \times10^{35}dyn\ cm^{-2}$, for all given values of central pressure ratio the mass of the NM sphere is larger than the DM sphere.
Therefore, the DMANSs are NM-dominated in these cases. However, for stars with ${p}_0 =0.4 \times10^{35}dyn\ cm^{-2}$ and $\delta\geq4$, as well as the stars with $m=1.0\ GeV$ and $\delta=5.00$, the DM
sphere is heavier than the NM one. Thus, these stars are DM-dominated.
It is obvious from Tables~\ref{table4}-\ref{table41} that for the neutron dominated DMANSs the halo of DM is confined within the
NM sphere. In contrast, in the DM dominated DMANSs the NM sphere is located inside the halo of DM which is in
agreement with other investigations \citep{Leung11,Leung12}.

\begin{figure*}
\includegraphics{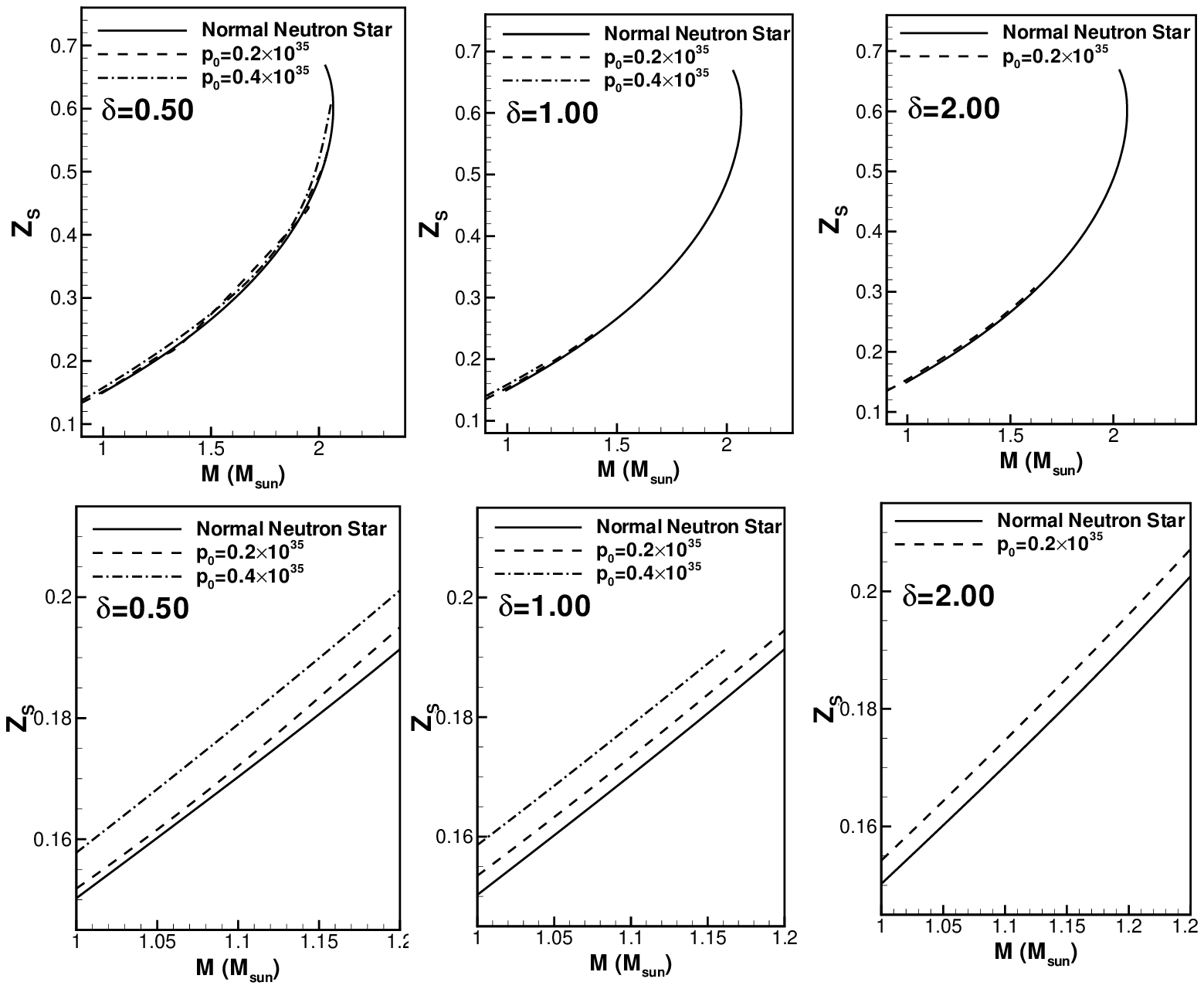}
\caption{Top: gravitational redshift, $Z_s$, versus total mass of DMANSs with different DMEOSs from the velocity profile of galaxies and central pressure ratio, $\delta$. Bottom: same as the top, but for a different range of DMANS mass.}
\label{fig10}
\end{figure*}
\begin{figure*}
\includegraphics{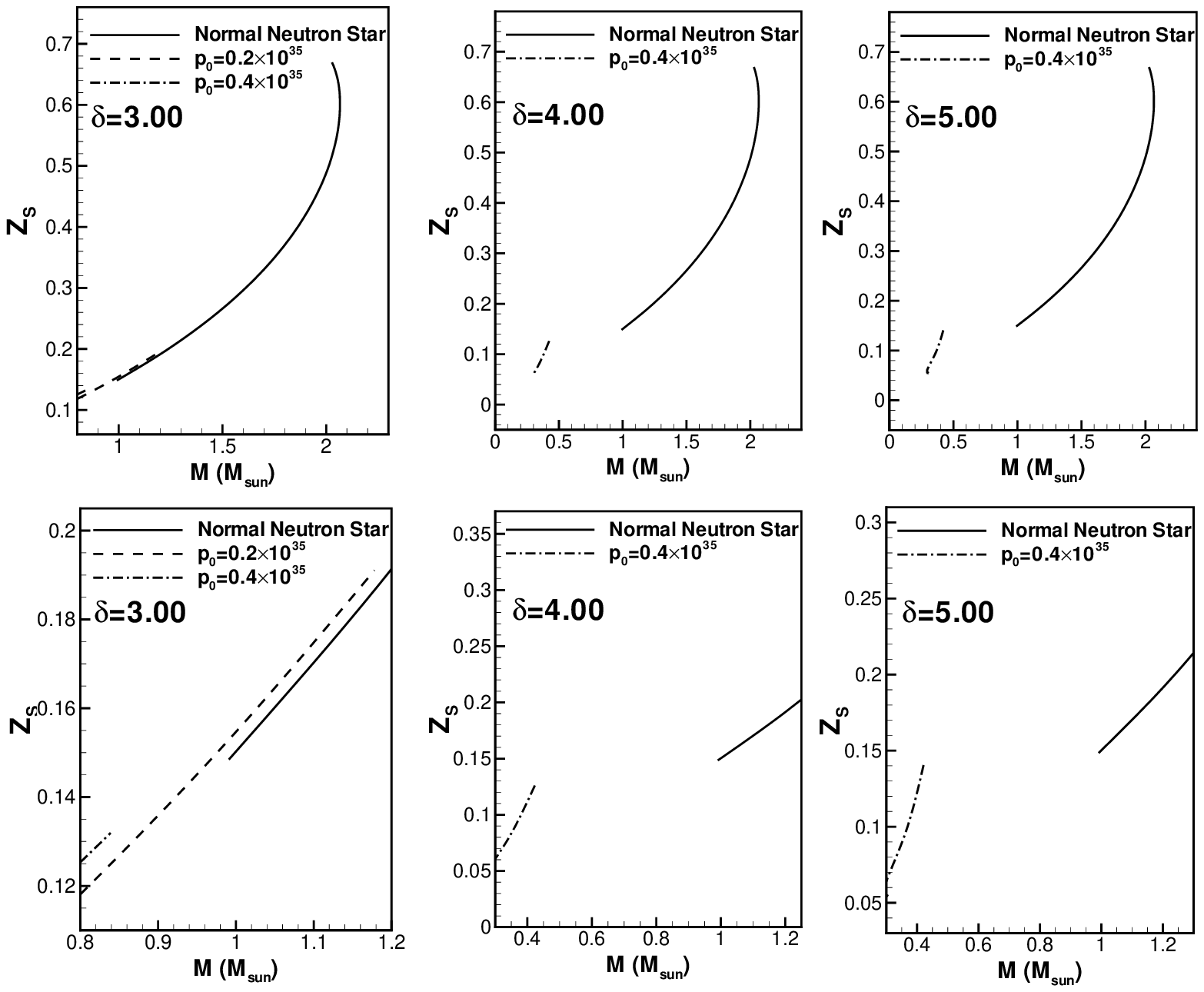}
\caption{Same as Figure \ref{fig10} but for other values of central pressure ratio, $\delta$.}
\label{fig11}
\end{figure*}

\begin{figure*}
\includegraphics{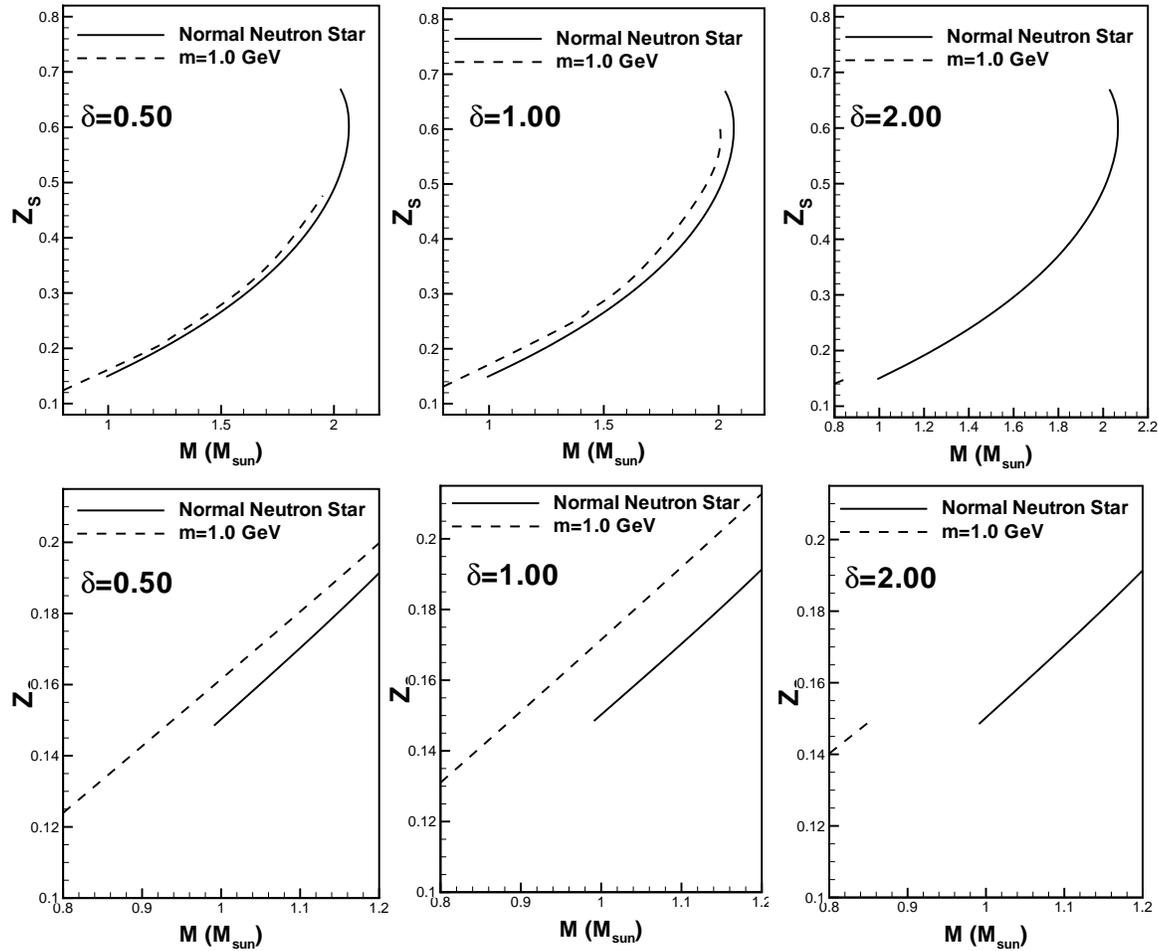}
\caption{Same as Figure \ref{fig10} but for DMANSs with the fermionic DMEOS for different central pressure ratios, $\delta$.}
\label{fig101}
\end{figure*}
\begin{figure*}
\includegraphics{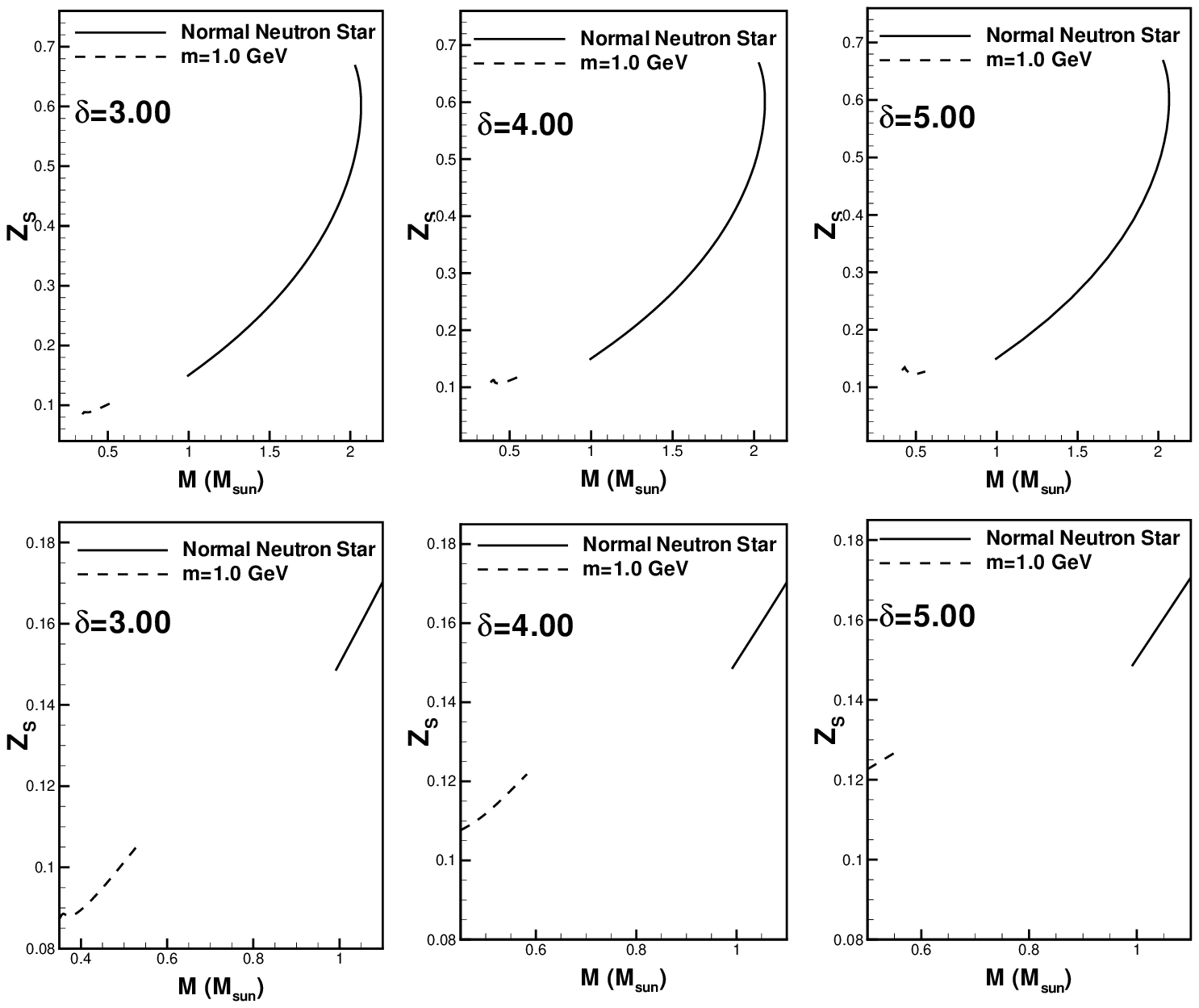}
\caption{Same as Figure \ref{fig101} but for other values of central pressure ratio, $\delta$.}
\label{fig111}
\end{figure*}

Considering two stars with the same mass-one of them a normal neutron star and the other
a DMANS in which the Nm sphere is surrounded by the halo of DM-they have the same
gravitational effects owing to their equal mass. However, the visible radius of the DMANS is smaller than the normal neutron star. Therefore, the gravitational redshift of spectral lines can be used to investigate these two stars.
Figures \ref{fig10}-\ref{fig111} present the gravitational redshifts of a normal neutron star and DMANSs with different DMEOSs, and various values of central pressure ratio. In all cases, the existence of DM leads to
higher values of gravitational redshift compared with normal neutron stars, in agreement with previous reports \citep{Leung11,Xiang}.
Moreover, the gravitational redshift in the DMANSs with $m=1.0\ GeV$ is larger than in the cases
with ${p}_0 =0.2$ and ${p}_0 =0.4\times10^{35}dyn\ cm^{-2}$. Figures \ref{fig10}-\ref{fig111} indicate that the effects of DM on the
gravitational redshift are more significant in stars with the higher values of
$\delta$. These results can be used to distinguish the normal neutron stars from DMANSs
with different DMEOSs and contributions of DM.


\section{Summary and Conclusions} \label{sec:highlight}

Applying the DMEOSs from the rotational curves of galaxies, as well as the fermionic DMEOS with $m=1.0\ GeV$ and
the NM EOS from Skyrme effective force, we have studied the
structure of DMANSs.
We have found that the radius of a DMANS is smaller than the radius of a normal neutron
star with the same mass. Our results show that the mass-radius relations are affected by both the DMEOS and central pressure ratio of DM to NM.
Specifically, depending on the DMEOS and the central pressure ratio, the DMANS
can be a gravitationally or self-bound star.
Additionally, it has been confirmed that in some cases two DMANSs with equal visible radii
but different masses can exist.
This fact is a consequence of different degrees of
DM domination in DMANSs.
We have shown that, with all DMEOSs, the maximum mass of the
DMANS is lower than the normal neutron star.
The observed neutron stars EXO 1745-248,
4U 1608-52, and 4U 1820-30,
which are inconsistent with the previous observations and theoretical properties
of normal neutron stars, can be explained with our model.
We have confirmed that some DMEOSs and central pressure ratios considered in this work result in DM-dominated neutron stars with a halo of DM around a small NM sphere.
Our calculations suggest that the observed neutron stars EXO 1745-248,
4U 1608-52, and 4U 1820-30
may be DM-dominated neutron stars.
It has been clarified that the radius of the NM sphere decreases
while the radius of the DM sphere
increases as the central pressure ratio grows.
In addition, the mass of the DM sphere increases with the increase
of central pressure ratio.
Moreover, the DMANSs with high values of central pressure ratio are
DM-dominated neutron stars.
Finally, we have shown that the existence of DM in the DMANSs results in increased gravitational redshift.

\acknowledgments
The author wishes to thank the Shiraz University Research Council and the Iran Science Elites Federation.



\allauthors

\listofchanges

\end{document}